\documentclass[
reprint,
floatfix,
 nofootinbib,
 amsmath,amssymb,
 aps,
 prb
]{revtex4-2}

\usepackage{graphicx}
\usepackage{dcolumn}
\usepackage{bm}
\usepackage{textcomp} 
\usepackage{mathtools} 
\usepackage{braket} 
\usepackage{color}
\usepackage{booktabs} 
\usepackage{multirow}

\def\l{\left}
\def\r{\right}
\def\nn{\nonumber}

\begin{document}


\title{Smooth velocity shuttling for suppressing valley excitations\\ 
in disordered Si/SiGe quantum dots}

\author{Ryo Nagai}
\email{ryo.nagai.jd@hitachi.com}
\affiliation{
Research and Development Group, 
Hitachi, Ltd., Kokubunji, Tokyo 185-8601, Japan
}

\author{Takashi Takemoto}
\affiliation{
Research and Development Group, 
Hitachi, Ltd., Kokubunji, Tokyo 185-8601, Japan
}

\author{Hiroyuki Mizuno}
\affiliation{
Research and Development Group, 
Hitachi, Ltd., Kokubunji, Tokyo 185-8601, Japan
}

\date{\today}

\begin{abstract}
Coherent electron shuttling is a key requirement for realizing scalable silicon quantum computing architectures. However, in silicon qubits, the existence of nearly degenerate conduction-band valleys poses a significant challenge because non-adiabatic transitions to excited valley states cause spin dephasing via spin-valley mixing. In this paper, we propose a \textit{smooth velocity shuttling} protocol to suppress these valley excitations. By mapping the time-domain design of the shuttling velocity profile onto the design problem of window functions in signal processing, we establish an analytical and intuitive design guideline that does not require computationally expensive numerical optimization. We demonstrate that the high-frequency sidelobes of the shuttling velocity spectrum can be effectively suppressed by applying a frequency-modulated gate voltage based on the Tukey window. Through statistical numerical simulations incorporating realistic spatial randomness of the valley landscape, we show that the proposed smooth velocity control significantly reduces the average spin infidelity in the moderate-to-low disorder regime ($|\Delta_0|/\sigma_\Delta \simeq \mathcal{O}(1)$). Our results underscore that this simple, control-level velocity shaping provides a robust pathway toward high-fidelity spin transport in large-scale silicon quantum processors.
\end{abstract}

\maketitle


\section{Introduction}
Quantum computers have attracted significant attention because they have the potential to achieve computational speeds that are difficult to reach with classical computers, thereby solving complex problems \cite{1994-Shor, 1996-Grover, 1996-Ekert-Jozsa, 1999-Shor}. However, current quantum devices have high error rates, making it difficult to fully benefit from quantum algorithms. Therefore, to unlock the true potential of quantum computing, the realization of fault-tolerant quantum computing implementing quantum error correction (QEC) is essential, and the development of technologies enabling high-fidelity and scalable quantum operations is strongly demanded \cite{1997-Kitaev-Yu, 2003-Kitaev-Yu, 2006-Raussendorf-et-al, 2007-Raussendorf-et-al, 2007-Raussendorf-2-et-al, 2011-Wang-et-al, 2012-Fowler-et-al, 2022-Tomaru-et-al}.

Although various physical systems such as superconducting circuits and ion traps are being researched and developed as implementations of quantum computers, this paper focuses on silicon quantum computers \cite{
1998-Loss-et-al, 
1999-Burkard-et-al, 
2013-Zwanenburg-et-al, 
2023-Burkard-et-al, 
2026-Mclntyre-Sarkar-Loss, 2026-HRL}. Silicon quantum computers, which utilize electrons in silicon as qubits, are expected to have long coherence times \cite{
2014-Veldhorst-et-al, 
2020-Struck-et-al, 
2022-Stano-et-al, 
2023-Tanttu-et-al, 
2024-Kuno-et-al, 
2025-Kuno-et-al} and extremely high compatibility with existing semiconductor nanofabrication technologies (CMOS processes) \cite{
2016-Maurand-et-al, 
2017-Veldhorst-et-al, 
2021-Xue-et-al, 
2020-Li-et-al, 
2021-Gonzalez-et-al, 
2022-Ruffino-et-al, 
2022-Zwerver-et-al, 
2023-Elsayed-et-al, 
2024-Neyens-et-al, 
2024-Stuyck-et-al, 
2026-HRL}. Consequently, they are drawing attention as one of the most promising platforms for realizing large-scale quantum processors that integrate a large number of qubits on a single chip.

To implement large-scale QEC in silicon quantum computers, an architecture capable of coupling many qubits under limited wiring resources is required. A crucial core technology that plays this role is ``electron shuttling'', which moves electron spins between quantum dots while preserving their coherence \cite{2019-Boter-et-al, 2022-Boter-et-al, 2023-Kunne-et-al, 2024-Siegel-et-al, 2025-Siegel-et-al, 2025-Yenilen-et-al, 2025-SpinHex, 2026-Moncy-et-al}. Driven by this motivation, high-fidelity shuttling techniques have been actively studied in recent years. Indeed, recent experiments have successfully demonstrated high-fidelity spin shuttling one after another \cite{
2021-Seidler-et-al,
2023-Xue-et-al,
2024-Struck-et-al,
2024-DeSmet-et-al,
2025-Matsumoto-et-al}.

Nevertheless, the shuttling required by truly scalable QEC architectures (such as the implementation of surface codes) is not limited to mere long-distance straight-line motion. More complex controls, such as accurate stopping at designated locations and changing directions at junctions like T-junctions, are required \cite{
2019-Boter-et-al, 2022-Boter-et-al, 2023-Kunne-et-al, 2024-Siegel-et-al, 2025-Siegel-et-al, 2025-Yenilen-et-al, 2025-SpinHex, 2026-Moncy-et-al}. In particular, when considering these starting and stopping motions, the ``discrete control of the shuttling velocity'' becomes a crucial factor as we will discuss below. However, conventional shuttling methods are not sufficiently equipped with the capability to continuously and smoothly control the shuttling velocity; in practice, their operations assume discrete velocity control or abrupt starting and stopping.

Such abrupt acceleration and deceleration without proper velocity control strongly induce non-adiabatic transitions, where the quantum state cannot follow the instantaneous eigenstates. In silicon qubits, because the energy splitting of the conduction-band valley degrees of freedom is extremely small, this non-adiabaticity easily causes transitions to high-energy excited valley states. Transitions to the excited valley states lead to subsequent irreversible relaxation accompanied by phonon emission and spin dephasing through valley-spin mixing, serving as a primary error source that destroys spin coherence \cite{2014-Kawakami-et-al,2015-Veldhorst-et-al,2018-Ferdous-et-al,2018-Ruskov-et-al,2023-Langrock-et-al, 2024-David-et-al, 2025-Volmer-et-al, 2026-Romero-et-al}.
Although recent theoretical studies have employed numerical optimization methods to find optimal velocity profiles that suppress these transitions \cite{2024-David-et-al, 2024-Oda-et-al, 2026-Romero-et-al}, they often operate as a black box. Consequently, translating such numerically generated, exotic velocity profiles into experimentally feasible gate-voltage waveforms remains a formidable challenge.

In this paper, we propose a new shuttling control scheme called \textit{smooth velocity shuttling} to fundamentally suppress such non-adiabatic transitions associated with abrupt acceleration and deceleration. We analytically show that the time-waveform design of the shuttling velocity is equivalent to the ``window function design problem'' in signal processing, and we propose a method to suppress sharp changes in the shuttling velocity by utilizing frequency-modulated (FM) gate voltages. Furthermore, through numerical simulations considering atomic-scale variations at the Si interface, we demonstrate that the proposed smooth velocity shaping effectively suppresses valley excitations and can serve as an important technology for achieving high-fidelity spin transport.
Furthermore, it is worth emphasizing that material-level advancements, such as the suppression of interface disorder, and our control-level velocity shaping strategy are not mutually exclusive but rather highly synergistic. By combining these two independent approaches, the robustness against valley variations can be significantly amplified, offering a highly effective and practical pathway toward the realization of fault-tolerant quantum computing architectures.

The structure of this paper is as follows. In Section 2, we introduce the theoretical framework of our proposed \textit{smooth velocity shuttling}. First, we discuss the relationship between valley excitations and the shuttling velocity waveform, and we propose a new shuttling protocol based on the Tukey window to suppress high-frequency components. We also present a specific implementation method using frequency-modulated (FM) gate voltages. In Section 3, we demonstrate the effectiveness of this approach through numerical simulations. In Section 3.1, we confirm through electrostatic potential calculations that continuous velocity control can be achieved via FM-modulated gate voltages. In Section 3.2, we evaluate the shuttling fidelity under the influence of spatial variations in valley coupling and clearly show that suppression of valley excitations and improvement in spin fidelity are achieved. Finally, in Section 4, we summarize the findings of this study and describe future outlooks\footnote{In this paper, we use natural units ($\hbar=c=1$) for simplicity, restoring physical units when necessary.}.

\section{Proposed Method}
\label{sec:Proposed Method}
In this section, we introduce the proposed method, \textit{smooth velocity shuttling}. First, in Sec.~\ref{sec:Valley excitation and shuttling velocity}, we introduce a simple effective model describing the valley states of electrons in silicon and discuss the relationship between valley excitations and the shuttling velocity. In the following Sec.~\ref{sec:Smooth velocity shuttling}, based on the considerations in Sec.~\ref{sec:Valley excitation and shuttling velocity}, we concretely introduce our proposed method, \textit{smooth velocity shuttling}. Furthermore, in Sec.~\ref{sec:Implementation}, we describe the implementation method of this technique.

\subsection{Valley excitation and shuttling velocity}
\label{sec:Valley excitation and shuttling velocity}
Let us consider the changes of the electronic state in Si associated with shuttling. In general, from the perspective of maintaining the quality of a qubit, state changes due to shuttling should be suppressed as much as possible. Therefore, a shuttling control design that suppresses excitations is extremely important.

An excitation that requires particular attention during electron shuttling in silicon is the excitation of the valley degree of freedom, which is unique to silicon \cite{2023-Langrock-et-al, 2025-Volmer-et-al}. The valley degree of freedom originates from the vicinity of the conduction band minima in silicon, and the energy gap between the ground state and the excited state is on the order of $\mathcal{O}(1-1000\,\mu\text{eV})$ \cite{2023-Degli-et-al,2024-Volmer-et-al,
2025-Marcks-et-al,
2025-Woods-et-al,
2026-HRL}. When using electrons in silicon as qubits, it is ideal to maintain the valley degree of freedom in the ground state. In the following, we focus on valley state transitions during shuttling.

Among the valley states in the silicon conduction band, we focus on two valley states corresponding to the Bloch states in the [001] direction, namely, the Bloch vectors ${\bm{k}}_+=(0,0,k_0)$ and ${\bm{k}}_-=(0,0,-k_0)$. Here, $k_0$ is a parameter related to the silicon lattice constant $a_0\simeq 0.53\,\text{nm}$, and specifically, it is given by $k_0=0.85\,(2\pi/a_0)\simeq (0.1\,\text{nm})^{-1}$. Choosing these valley states $\ket{{\bm{k}}_+}$ and $\ket{{\bm{k}}_-}$ as the basis, the valley Hamiltonian $H_{v0}$ for a quantum dot at rest at position ${\bm{r}}_{\text{QD}}$ can be expressed using the complex valley coupling $\Delta({\bm{r}}_{\text{QD}})=|\Delta({\bm{r}}_{\text{QD}})|e^{i\phi_v({\bm{r}}_{\text{QD}})}$ as follows \cite{2006-Friesen-Eriksson-Coppersmith,2007-Friesen-et-al}\footnote{This is derived by using the effective mass approximation and considering perturbations against the breaking of crystal symmetry \cite{2006-Friesen-Eriksson-Coppersmith,2007-Friesen-et-al}. The Hamiltonian (\ref{eq:H_v0}) is the result incorporating up to the first-order perturbation. Note that as a result of the first-order perturbation, a common energy factor appears in the diagonal terms, but its contribution has been removed by a global phase transformation.}:
\begin{align}
H_{v0} 
&= 
\begin{pmatrix} 
0 & \Delta({\bm{r}}_{\text{QD}}) \\ 
\Delta^*({\bm{r}}_{\text{QD}}) & 0 
\end{pmatrix} 
\nn\\
&= 
\frac{E_v({\bm{r}}_{\text{QD}})}{2} \left[
\tau_x\cos\phi_v({\bm{r}}_{\text{QD}}) 
- 
\tau_y\sin\phi_v({\bm{r}}_{\text{QD}}) 
\right]
\,.
\label{eq:H_v0}
\end{align}
Here, $E_v({\bm{r}}) = 2|\Delta({\bm{r}})|$ is the local valley splitting energy at position ${\bm{r}}$, $\phi_v({\bm{r}}) = \text{arg}(\Delta({\bm{r}}))$ is the valley phase, and $\tau_x, \tau_y$ are the Pauli matrices defined in the space spanned by the valley states $\ket{{\bm{k}}_\pm}$ of the stationary system.

When an electron moves according to the shuttling trajectory ${\bm{r}}_{\text{QD}}(t)\equiv {\bm{r}}_{\text{QD}}$, the valley Hamiltonian in the stationary system becomes time-dependent through the form $H_{v0}(t) = H_{v0}({\bm{r}}_{\text{QD}})$. At this time, to make the quantum state follow the valley phase $\phi_v({\bm{r}}_{\text{QD}})$ which varies with position, we introduce a unitary transformation to a rotating coordinate system based on the instantaneous eigenstates, $U_v(t) = \exp(-i \frac{\phi_v({\bm{r}}_{\text{QD}})}{2} \tau_z)\,\mathtt{H}$, where $\mathtt{H}$ represents the Hadamard transformation in $\ket{{\bm{k}}_\pm}$ basis. Calculating the effective Hamiltonian after the transformation, $H_{v}(t) = U^\dagger_v H_{v0} U_v - i U^\dagger_v \dot{U}_v$, yields the following equation \cite{2025-Nagai-et-al,2025-QTCAD,2024-David-et-al, 2026-Romero-et-al}:
\begin{align}
H_{v}(t) 
&= 
\frac{1}{2}
\begin{pmatrix} 
E_v({\bm{r}}_{\text{QD}}) &
\dot{\phi}_v({\bm{r}}_{\text{QD}}) \\ 
\dot{\phi}_v({\bm{r}}_{\text{QD}}) &
-E_v({\bm{r}}_{\text{QD}}) 
\end{pmatrix} 
\,\nn\\
&=
\frac{1}{2} 
\left[
E_v({\bm{r}}_{\text{QD}}) 
\tilde{\tau}_z({\bm{r}}_{\text{QD}}) 
+ 
\dot{\phi}_v({\bm{r}}_{\text{QD}}) 
\tilde{\tau}_x ({\bm{r}}_{\text{QD}}) 
\right]
\,,
\label{eq:H_v}
\end{align}
where the Pauli matrices $\tilde{\tau}_z({\bm{r}}_{\text{QD}}) , \tilde{\tau}_x({\bm{r}}_{\text{QD}}) $ are defined in the space spanned by the local valley ground state $\ket{g_v({\bm{r}}_{\text{QD}})  }=(\ket{k_+}-e^{-i\phi_v({\bm{r}}_{\text{QD}})}\ket{{\bm{k}}_-})/\sqrt{2}$ and the excited state $\ket{e_v({\bm{r}}_{\text{QD}})  }=(\ket{k_+}+e^{-i\phi_v({\bm{r}}_{\text{QD}})}\ket{{\bm{k}}_-})/\sqrt{2}$ in the rotating coordinate system. From this equation, the factor causing valley state transitions during shuttling is identified: it is the off-diagonal component $\dot{\phi}_v({\bm{r}}_{\text{QD}})$. Paying attention to the chain rule $\dot{\phi}_v({\bm{r}}_{\text{QD}}) = \dot{\bm{r}}_{\text{QD}}(t) \cdot {\bm{\nabla}} \phi_v = {\bm{v}}(t) \cdot {\bm{\nabla}} \phi_v$, this is directly related to the shuttling velocity ${\bm{v}}(t)=\dot{\bm{x}}_{\text{QD}}(t)$. Therefore, we see that appropriately controlling the shuttling velocity is crucial to suppressing valley excitation transitions.

The specific relationship between valley excitation transitions and the shuttling velocity can be obtained by solving the time-dependent Schrödinger equation based on the Hamiltonian (\ref{eq:H_v}). Assuming $\|\dot{\phi}_v/E_v\|\ll1$ and using the perturbation approximation\footnote{The validity of this assumption will be confirmed in subsequent numerical calculations.}, the probability $p_v(t_f,t_i)$ that a state which was in the ground state at the shuttling start time $t=t_i$ is found in the excited valley state at time $t_f\,(\ge t_i)$ is given by
\begin{align}
p_v(t_f,t_i)
\,=\,
\biggl\|
\int^{t_f}_{t_i}
\left(
\frac{{\bm{v}}\cdot{\nabla \phi_v}}{2}
\right)
e^{i E_v t}
dt
\biggr\|^2
\,,
\label{eq:pv-gen}
\end{align}
Therefore, if we simply assume that the valley coupling is uniform along the shuttling path, meaning that $E_v$ and ${\bm{\nabla}}\phi_v$ do not depend on time $t$\footnote{As described later, this assumption is unrealistic, and in actual silicon quantum devices, it becomes non-uniform due to variations at the interface. The effects of these non-uniformities will be confirmed by subsequent numerical calculations.}, Eq.~(\ref{eq:pv-gen}) implies that the valley state excitation rate $p_v$ after the completion of shuttling is given by
\begin{align}
p_v
\,=\,
\biggl\|
\left(
\frac{{\nabla \phi_v}}{2}
\right)
\cdot
\int^{\infty}_{-\infty}
{\bm{v}}\,
e^{i E_v t}
dt
\biggr\|^2
\,=\,
\biggl\|
\left(
\frac{{\nabla \phi_v}}{2}
\right)
\cdot
\tilde{\bm{v}}(\omega_v)
\,
\biggr\|^2
\,.
\label{eq:pv_end}
\end{align}
where $\omega_v=E_v$. In the first equality of Eq.~(\ref{eq:pv_end}), we used the fact that the shuttling velocity has non-zero values only for $t_i\leq t\leq t_f$, and in the second equality of Eq.~(\ref{eq:pv_end}), we introduced the Fourier transform of ${\bm{v}}(t)$
\begin{align}
\tilde{\bm{v}}(\omega)
\,=\,
\int^\infty
_{-\infty}
dt\,
{\bm{v}}(t)
e^{-i\omega t}
\,,
\end{align}
and further utilized the symmetry with respect to the origin $\tilde{\bm{v}}(\omega)=\tilde{\bm{v}}(-\omega)$. From the above Eq.~(\ref{eq:pv_end}), it is found that the valley excitation transition rate is determined by the $\omega=\omega_v=E_v$ component in the spectrum of the velocity signal ${\bm{v}}(t)$. Although $\omega_v$ depends on the vertical structure of the device and the shape of the quantum dot, it is typically $\omega_v\simeq\mathcal{O}(\text{GHz})$.

Based on the above argument, the optimal velocity design can be reduced to a signal processing problem: ``constructing a velocity waveform ${\bm{v}}(t)$ that suppresses a specific frequency component $|\tilde{\bm{v}}(\omega_v)|$.'' Regarding the velocity waveform ${\bm{v}}(t)$ as a window function with a finite time width, its spectrum is characterized by a mainlobe and sidelobes. 
Focusing on the frequency scales involved, the argument can be summarized as follows:
First, the total shuttling time $T=(t_f-t_i)\simeq \mathcal{O}(10\,\mu\text{s})$ determines the width of the mainlobe, which is given by $1/T=\mathcal{O}(10^{-4}\,\text{GHz})$. 
Second, the high-frequency component that actually contributes to the valley excitation rate is on the order of $\omega_v \simeq \mathcal{O}(\text{GHz})$. 
Third, because the target frequency $\omega_v$ is several orders of magnitude higher than the mainlobe width, any broadening of the mainlobe---which is a typical trade-off in window function design---has a negligible impact on the excitation probability. 
Therefore, the essential challenge in our velocity design is exclusively concentrated on suppressing the high-frequency sidelobes. Consequently, we establish ``a velocity waveform that suppresses sidelobes'' as our core strategy for controlling the shuttling velocity.

Finally, we mention the assumptions introduced to simplify the discussion in this section. First, although we focused only on valley excitations in this section, to estimate the impact on the quality (fidelity) of silicon qubits, it is necessary to evaluate the changes in the spin state accompanying this valley excitation \cite{2014-Kawakami-et-al,2015-Veldhorst-et-al,2018-Ferdous-et-al,
2018-Ruskov-et-al,2023-Langrock-et-al, 2024-David-et-al, 2026-Romero-et-al}. These evaluations will be carried out in the subsequent Sec.~\ref{sec:Numerical Simulation}. Next, regarding the uniformity of valley coupling, although we ignored the variation of valley coupling for simplicity in the discussion after Eq.~(\ref{eq:pv_end}), valley coupling is non-uniform in actual devices \cite{2022-Wuetz-et-al,2023-Losert-et-al,2023-Degli-et-al,2024-Volmer-et-al,2025-Volmer-et-al,2025-Woods-et-al,2026-HRL}. The impact of this non-uniformity will be evaluated by numerical calculations in the subsequent Sec.~\ref{sec:Numerical Simulation}. Both assumptions were introduced to simplify the discussion, but subsequent numerical calculations will reveal that our proposed policy of a ``shuttling velocity that suppresses sidelobes'' fulfills an essential function in suppressing fidelity degradation associated with shuttling.

\subsection{Smooth velocity shuttling}
\label{sec:Smooth velocity shuttling}

\begin{figure}[t]
 \centering
 \includegraphics[width=8cm]{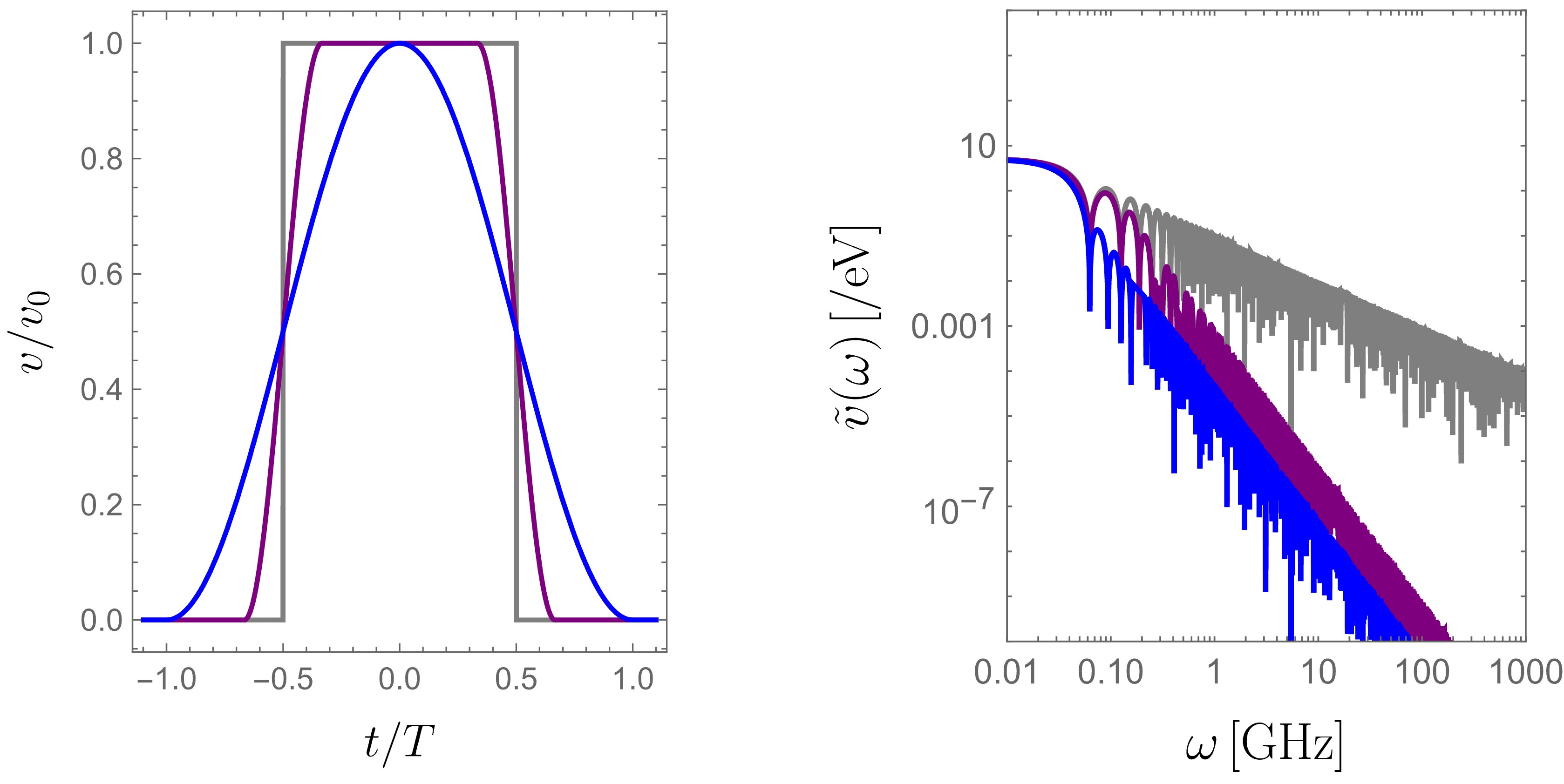}
 \caption{Shuttling velocity profiles (left) and spectra (right). The black curve corresponds to the time-dependence of the conventional shuttling velocity control ($\beta=0$), and the light-blue and blue curves correspond to that of our proposed shuttling controls with $\beta=0.5$ and $\beta=1$, respectively.}
 \label{fig:Tukey}
\end{figure}
Let us next introduce our idea on the shuttling velocity control.
In this subsection, we focus on the shuttling velocity 
along a specific spatial direction. To distinguish this  
analysis from the general multidimensional framework discussed in the previous section, the velocity is denoted by the lowercase scalar $v(t)$. It should be noted that for multidirectional shuttling scenarios, the same theoretical formulation can be applied independently to each spatial component.

Conventional shuttling control is designed assuming discrete shuttling velocity control, as shown by the black line in the right panel of Fig.~\ref{fig:Tukey}. That is, the basic waveform of the shuttling velocity is a rectangular window. In this case, the Fourier transform of the velocity waveform is as shown by the black line in the left panel of Fig.~\ref{fig:Tukey}, leaving many high-frequency components. As mentioned 
in Sec.~\ref{sec:Valley excitation and shuttling velocity}, this becomes a factor that induces valley excitations.

Therefore, in this study, as a shuttling velocity control that removes high-frequency components inducing valley excitations, we propose continuous shuttling velocity control (\textit{Smooth velocity shuttling}) based on a continuous waveform as shown by the purple or blue lines in the right panel of Fig.~\ref{fig:Tukey}, instead of the conventional rectangular wave control. Specifically, we employ a Tukey-window-based velocity waveform with the maximum velocity fixed at $v_0$ and the total travel distance (the integral of velocity) fixed at $v_0T$:
\begin{widetext}
\begin{align}
v(t)\,=\,
 \begin{dcases*}
    0
& if $t<-\frac{T'}{2}$, \\
   {\frac{v_0}{2}\biggl[
   1+\cos\l(\frac{2\pi}{\beta T'}\l(t+\frac{1-\beta}{2}T'\r)\r)
   \biggr]} & if $-T'/2\leq t<-\frac{1-\beta}{2}T'$, \\
   v_0 & if $-\frac{1-\beta}{2}T'\leq t \leq \frac{1-\beta}{2}T'$, \\
   {\frac{v_0}{2}\biggl[
   1+\cos\l(\frac{2\pi}{\beta T'}\l(t-\frac{1-\beta}{2}T'\r)\r)
   \biggr]} & if $\frac{1-\beta}{2}T'< t\leq \frac{T'}{2}$, \\
    0
& if $\frac{T'}{2}< t$,
 \end{dcases*}
 \,.
 \label{eq:v}
\end{align}
\end{widetext}
as the basic waveform of the shuttling velocity. Here, $T'$ is defined by
\begin{align}
T'\,=\,
\frac{T}{1-\beta/2}
\,,
\label{eq:T-prime}
\end{align}
which determines effective shuttling time in this scheme.
$\beta$ is a shape parameter that regulates the smoothness of the velocity waveform and is a real number satisfying $0\leq \beta\leq 1$. $\beta=0$ corresponds to the rectangular window equivalent to the conventional method, and $\beta=1$ corresponds to the smoothest waveform (Hann window). Waveforms for the examples of $\beta=0,0.5,1$ are shown on the left of Fig.~\ref{fig:Tukey}. The black line corresponds to the rectangular window ($\beta=0$), the purple line to $\beta=0.5$, and the blue line to $\beta=1$. By setting $\beta>0$, it can be seen that the velocity is continuously adjusted and the acceleration is smoothly shaped.

The most significant feature of the \textit{smooth velocity shuttling} proposed in this study is that the sidelobes of the velocity spectrum are strongly suppressed. Due to this property, an effect of suppressing valley excitations during shuttling is expected. Indeed, the Fourier transform $\tilde{v}(\omega)$ of the shuttling velocity waveform (\ref{eq:v}) proposed in this study is given by
\begin{align}
\tilde{v}(\omega)
\,=\,
\frac{2v_0}{\omega}
\sin\l(\frac{\omega T}{2}\r)
\mathcal{F}(\beta,\omega,T)
\,,
\label{eq:tilde-v}
\end{align}
\begin{align}
\mathcal{F}(\beta,\omega,T)
\,=\,
\frac{\pi^2\cos\l(\frac{1}{2}\frac{\beta\omega T}{2-\beta}\r)}{\pi^2-\l(\frac{\beta\omega T}{2-\beta}\r)^2}
\,,
\end{align}
and its shape is shown by the purple and blue lines in the right panel of Fig.~\ref{fig:Tukey}. 
Combining Eqs.~(\ref{eq:pv_end}) and (\ref{eq:tilde-v}), we find that,
compared to the spectrum of conventional shuttling velocity control (black line), high-frequency components that cause valley excitations are strongly suppressed, and it can be seen that the degree of suppression is controlled by the smoothness parameter $\beta$ of the velocity waveform. Therefore, by choosing an appropriately large $\beta$, it is expected that $|\tilde{v}(\omega_v)|$ can be reduced, thereby suppressing the excitation rate according to Eq.~(\ref{eq:pv_end}).

\subsection{Implementation}
\label{sec:Implementation}

\begin{figure}[t]
 \centering
 \includegraphics[width=8cm]{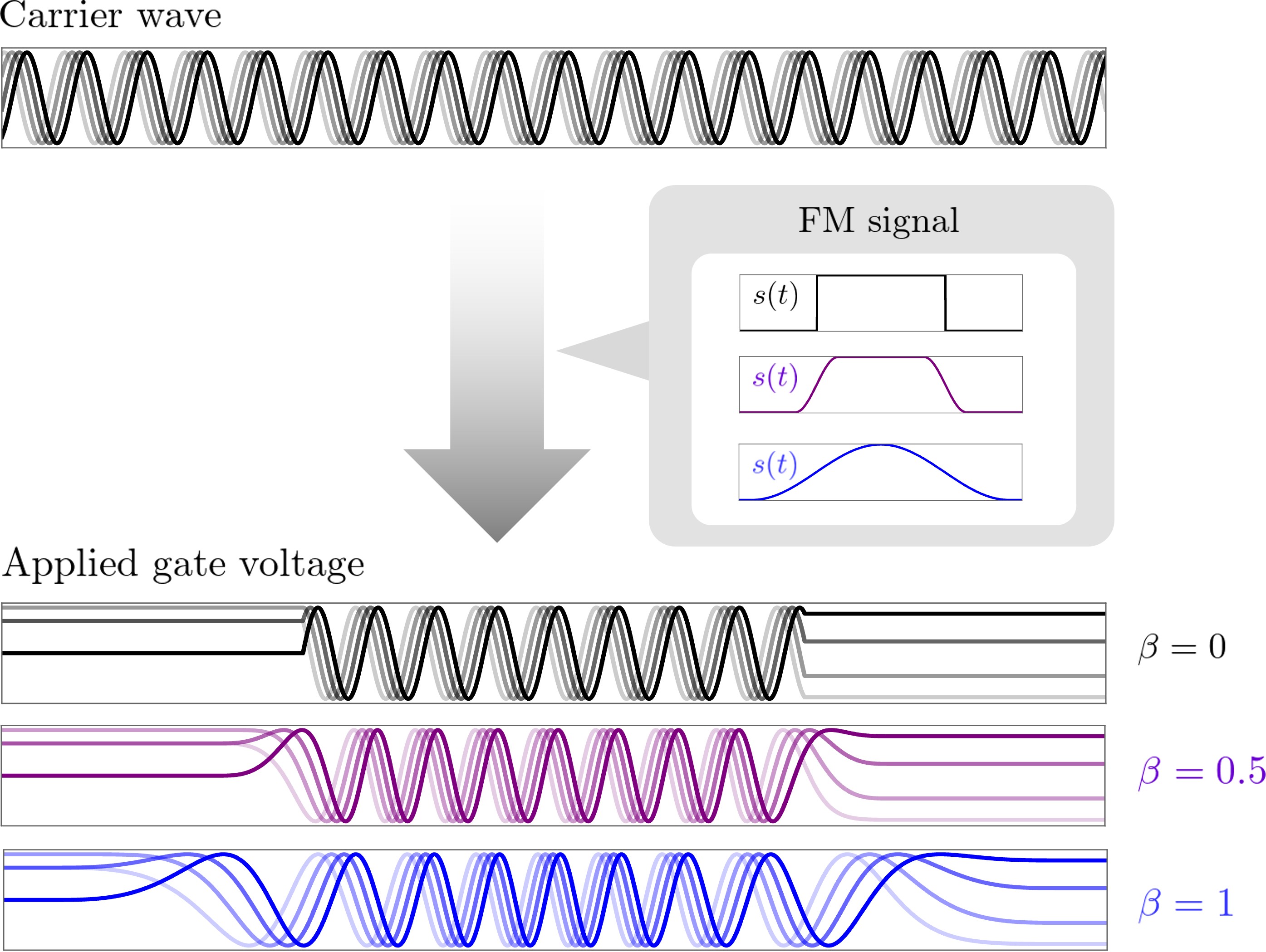}
 \caption{Gate-voltage waveforms for implementing the proposed shuttling method ({\textit{smooth velocity shuttling}}). 
 The gate-voltage waveforms (\ref{eq:V}) can be generated by frequency modulation of the conventional sine waveforms (\ref{eq:V-const}) by signal $s(t)$. The time dependence of the waveforms are shown in the top, middle, and bottom panels corresponding to $\beta=0$, $0.5$, and $1$, respectively.}
 \label{fig:Tukey-gate-voltage}
\end{figure}

\begin{figure*}[t]
\centering
\includegraphics[width=10cm]{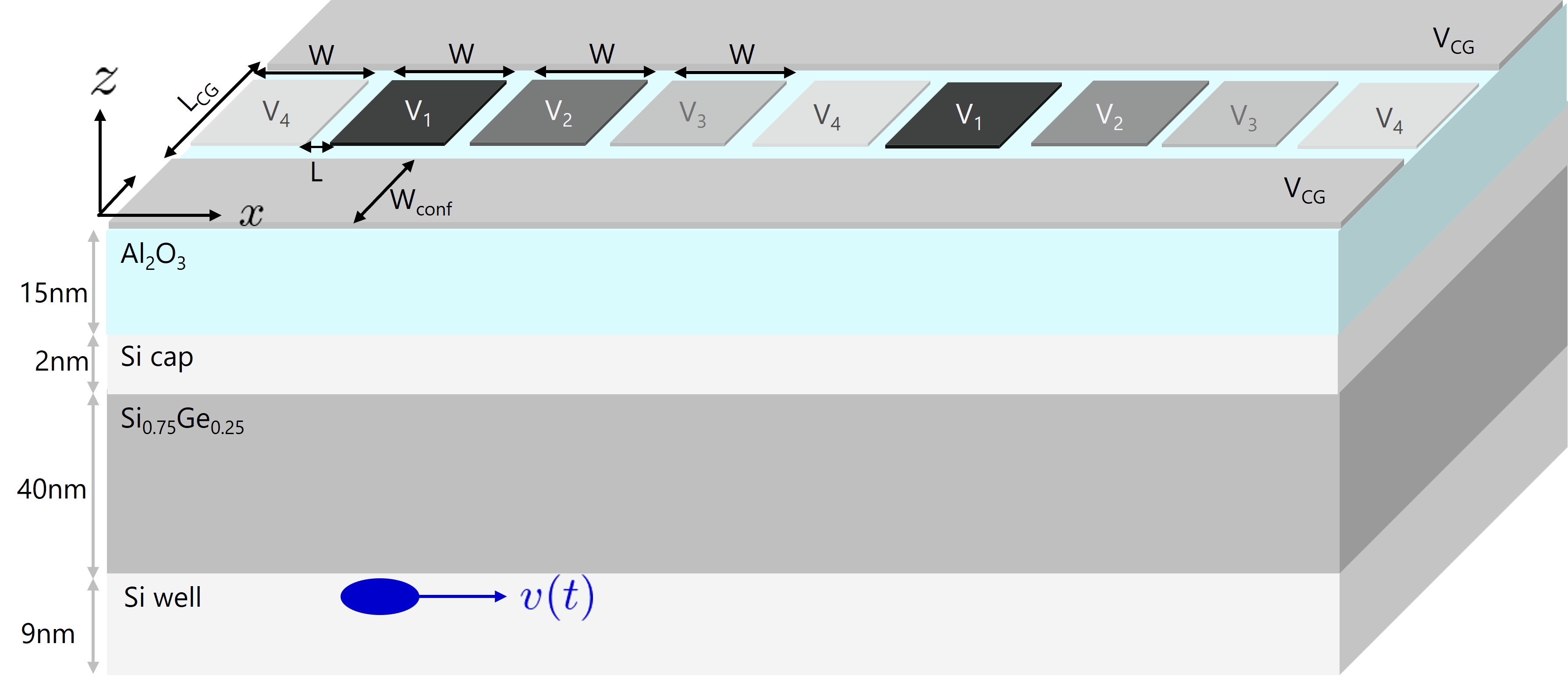}
\caption{Schematic diagram of the SiGe/Si/SiGe quantum dot shuttling device used in the numerical simulations. Gate electrodes are arranged at intervals of $L$ along the shuttling direction ($x$-axis), and a frequency-modulated (FM) voltage $V_m(t)$ with a period of $N=4$ is applied. Rectangular gates for confining electrons in the $y$-direction are placed on both sides of the shuttling path, and a constant DC voltage $V_{\text{CG}}$ is applied. Electrons are confined within the quantum dots formed by these electrodes and transported through the Si well.}
\label{fig:device}
\end{figure*}

\begin{figure*}[t]
 \centering
 \includegraphics[width=15cm]{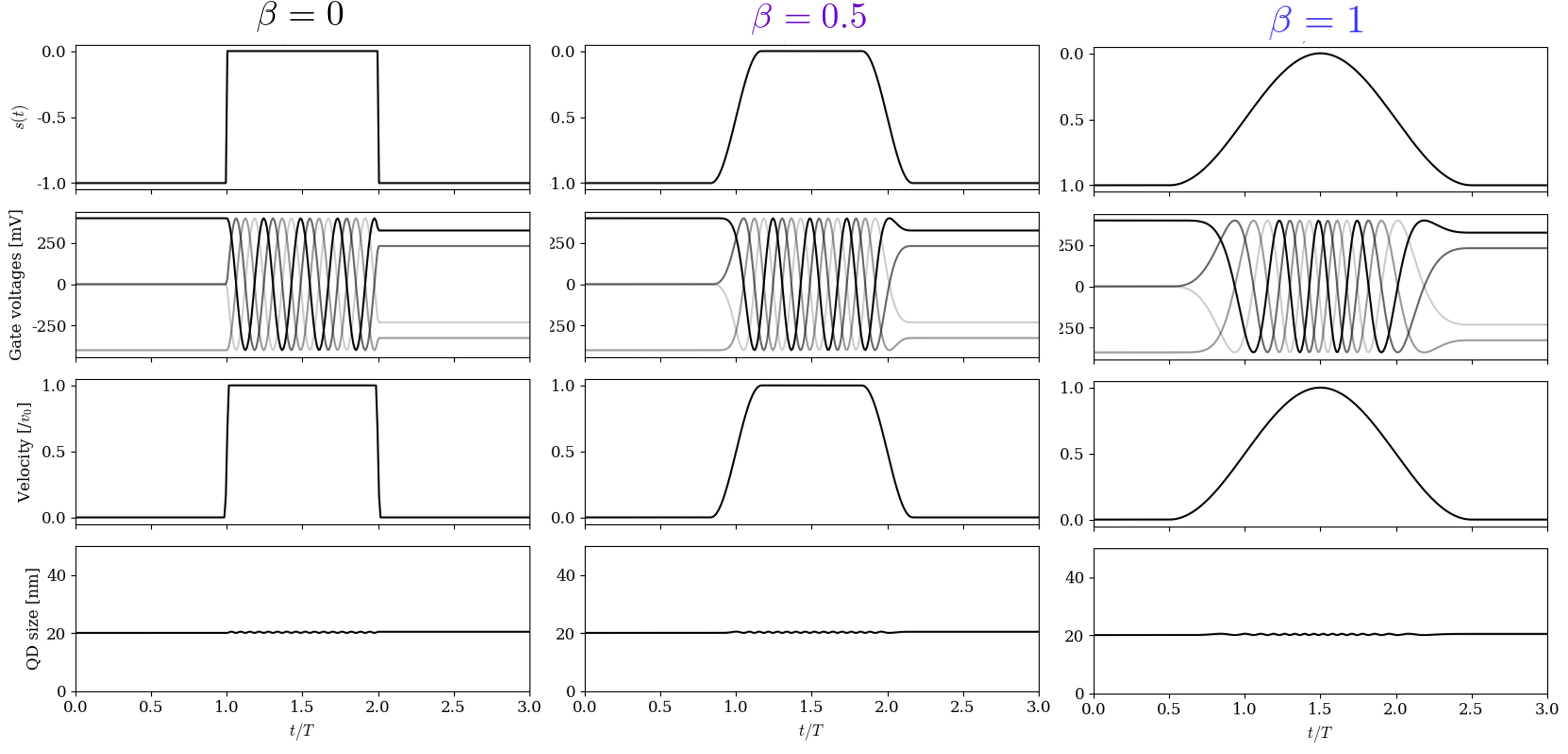}
 \caption{Numerical calculation results of the smooth velocity shuttling protocol. From top to bottom, the changes of the signal wave $s(t)$, the applied gate voltage $V_n(t)$, the shuttling velocity $v(t)$, and the quantum dot size $a_x(t)$ against the normalized time $t/T$ are shown. It can be confirmed that the acceleration during transport is modulated by the velocity shaping parameter $\beta$, and the high-frequency components are suppressed.}
 \label{fig:potential-sim}
\end{figure*}

In the previous section, we introduced \textit{smooth velocity shuttling}, which continuously controls the shuttling velocity waveform. In this section, we describe its specific implementation method.

In general, electron shuttling in silicon is implemented by applying periodic gate voltages to gate electrodes arranged along the shuttling direction. Below, we first review the conventional constant-velocity shuttling implementation method to clarify the relationship between the shuttling velocity and the applied voltage. Then, based on that relationship, we show the implementation method of the proposed method ({\textit{smooth velocity shuttling}}). Note that for simplicity, we consider one-dimensional shuttling below, but it can be similarly extended to multiple dimensions.

First, we describe the conventional shuttling velocity control. Conventional shuttling is implemented by arranging $N$ electrodes as one period along the shuttling direction and applying a time-dependent voltage given by
\begin{align}
V_n(t)
\,=\,
V_0\cos\l(
\omega_0 t
+
\phi_n
\r)
+
B_n
\,,
\label{eq:V-const}
\end{align}
to the $n$-th gate electrode \cite{
2021-Seidler-et-al,
2023-Langrock-et-al,
2023-Kunne-et-al,
2023-Xue-et-al,
2024-Struck-et-al,
2024-Volmer-et-al,
2024-DeSmet-et-al,
2024-Nemeth-et-al,
2025-Volmer-et-al,
2025-QTCAD,
2026-Beer-et-al,
2026-Undseth-et-al,
2026-Volmer-et-al}. Here, $V_0, \omega_0, \phi_n, B_n$ are the amplitude, angular frequency, phase, and DC bias of the applied voltage, respectively. $V_0$ acts as a parameter to adjust the quantum dot size, and is typically $V_0\simeq\mathcal{O}(100\,\text{mV})$. $\phi_n$ and $B_n$ are used to stabilize the shuttling trajectory. In this scheme, the shuttling velocity is controlled by the frequency $\omega_0$, and the velocity $v_0$ and the frequency $\omega_0$ satisfy the proportional relationship
\begin{align}
v_0=\frac{L\omega_0}{2\pi}
\,.
\label{eq:v0-Lf0}
\end{align}
Here, $L$ represents the travel distance per period. In the conventional method, since the frequency $\omega_0$ corresponding to the shuttling velocity is a constant value, it does not possess the capability to continuously adjust the shuttling velocity. Therefore, when including the starting and stopping operations, it results in a rectangular velocity waveform as shown by the black line in the left panel of Fig.~\ref{fig:Tukey}.

Next, we consider the implementation of \textit{smooth velocity shuttling}, which is the method proposed in this study. The core idea is to introduce frequency modulation into the gate voltage to vary the shuttling velocity over time. Specifically, for the $n$-th gate electrode, a voltage of
\begin{align}
V_n(t)
\,=\,
V_0\cos\l(
\omega_0t
+
\phi_n
+
\omega_0
\int^t d\tau
s(\tau)
\r)
+
B_n
\,,
\label{eq:V}
\end{align}
is applied. Here, $s(t)$ is given by
\begin{align}
s(t)
=
\frac{v(t)-v_0}{v_0}
\,,
\end{align}
and the desired velocity waveform, Eq.~(\ref{eq:v}), is substituted into $v(t)$. Equation (\ref{eq:V}) means that this method can be implemented by using the waveform of the conventional method (\ref{eq:V-const}) as a carrier wave and applying a voltage waveform frequency-modulated (FM) by the signal wave $s(t)$ to each gate electrode. The schematic figure for the implementation is shown on Fig. \ref{fig:Tukey-gate-voltage}.

A comparison of the applied voltage waveforms between the conventional method and the proposed method is shown on the bottom figures of Fig.~\ref{fig:Tukey-gate-voltage}. Examples for $\beta=0,0.5$ and $1$ are shown from top to bottom. Here, $N=4$ was assumed. The contrast of colors corresponds to the difference in electrode numbers. In the conventional method ($\beta=0$), the waveform applied to each electrode is a sine wave of a single frequency and is filtered by step signal function, whereas in the proposed method, it becomes a waveform modulated in the time direction by frequency modulation. Due to this frequency modulation, the shuttling velocity exhibits a continuous change as shown by the purple and blue lines in the left panel of Fig.~\ref{fig:Tukey}.

Finally, we mention the differences from previous studies on shuttling velocity control. Previous literatures \cite{2024-David-et-al, 2024-Oda-et-al, 2026-Romero-et-al} also discuss optimal shuttling velocities that extend conventional constant-velocity control to suppress fidelity degradation. However, because these methods determine the optimal velocity based on numerical optimization, it is not easy to analytically find the applied voltage waveforms required to realize such shuttling. In contrast, in this study, we analyze the non-adiabatic valley excitations accompanying shuttling from the perspective of signal processing as the ``sidelobes of the Fourier spectrum of the velocity profile,'' and propose a method to shape and control the velocity via frequency modulation using the Tukey window. This method has the advantage of being able to improve fidelity with a simple implementation while avoiding complex pulse shaping associated with numerical optimization.

\section{Numerical Simulation}
\label{sec:Numerical Simulation}
In the previous section, based on a simplified physical model, we proposed a new method called \textit{smooth velocity shuttling} to suppress the degradation of shuttling fidelity. In this section, we extend the model to reflect realistic situations and clarify the effectiveness of the method through numerical verification. Specifically, first, in Sec.~\ref{sec:potential simulation}, we verify that \textit{smooth velocity shuttling} can actually be realized by the implementation method introduced in Sec.~\ref{sec:Implementation}. Subsequently, in Sec.~\ref{sec:fidelity simulation}, based on a physical model reflecting a realistic situation, we numerically evaluate the effectiveness of the proposed method.

\subsection{Smooth Velocity Shuttling via FM modulated gate voltages}
\label{sec:potential simulation}
In this section, we estimate the electrostatic potential (quantum dot shape) and numerically verify that \textit{smooth velocity shuttling} is achieved under the voltage conditions of Eq.~(\ref{eq:V}).

As an example, let us consider a SiGe/Si/SiGe device as shown in Fig.~\ref{fig:device}. We assume that gates with a period of $N=4$ are arranged at intervals of $L$ along the shuttling direction ($x$-axis direction), and a voltage of $V_{m}(t)\,(m=(n - 1) \bmod 4 + 1)$ is applied to the $n$-th gate. We use Eq.~(\ref{eq:V}) as the specific form of the voltage, and $v(t)$ included in the frequency modulation is given by Eq.~(\ref{eq:v}). Regarding the gate shape, for simplicity, the thickness of the gate ($z$ direction) is ignored, and the shape is assumed to be a square of length $W$. Furthermore, to shape the quantum dot geometry, in addition to the above gates, new rectangular gates are added. A time-independent constant voltage $V_{\text{CG}}$ is assumed to be applied to these gates.

Evaluating the shape of the quantum dot $V_{\text{QD}}({\bm{r}})$ in silicon corresponds to estimating the electrostatic potential $\phi({\bm{r}})$ formed by the above voltages. The electrostatic potential $\phi({\bm{r}})$ is found by solving the Laplace equation ${\bm{\nabla}}^2\phi({\bm{r}})=0$ using the device structure and voltage conditions as boundary conditions\footnote{To include the effects of impurities in the device, it is necessary to solve the Poisson equation containing the charge source term. Here, for simplicity, their contributions are assumed to be negligible.}. In the present case, since the gate voltage depends on time $t$ according to Eq.~(\ref{eq:V}), the electrostatic potential is also time-dependent. Although the solution to the Laplace equation ${\bm{\nabla}}^2\phi({\bm{r}})=0$ depends on the gate shape, it is known that it can be found analytically for the rectangular shape introduced above \cite{1995-Davies-et-al} Although the solution to the Laplace equation ${\bm{\nabla}}^2\phi({\bm{r}})=0$ depends on the gate shape, it is known that it can be found analytically for the rectangular shape introduced above \cite{1995-Davies-et-al, 2026-Kanno-et-al}{\footnote{While the original analytical result by Davies \textit{et al.} \cite{1995-Davies-et-al} considers a single-layer gate structure along the $z$-direction, we extended the formulation to a multi-layer heterostructure to incorporate the realistic vertical layout of our device.}}. In the following analysis, we use this analytical expression to calculate $\phi({\bm{r}})$ and determine the quantum dot shape $V_{\text{QD}}({\bm{r}})=-e\phi({\bm{r}})$ therefrom. Here, $e$ represents the elementary charge. When evaluating the electrostatic potential, we assumed that a quantum dot is formed immediately below the SiGe and Si well interface at $2\,\text{nm}$, and found the electrostatic potential $\phi({\bm{r}})=\phi(x,y,z_{\text{QD}})$ at $z=z_{\text{QD}}=15\,\text{nm}+2\,\text{nm}+40\,\text{nm}+2\,\text{nm}$. In addition, when determining the electrostatic potential, the relative permittivities of $\text{Al}_2\text{O}_3$, Si, and SiGe were set to 9, 13.2, and 11.8, respectively.

Figure \ref{fig:potential-sim} shows the results of the time dependence of the electrostatic potential (quantum dot shape). In this analysis, the device structure parameters were set to $W=60\,\text{nm}$, $L=1\,\text{nm}$, $W_{\text{conf}}=60\,\text{nm}$, $L_{\text{CG}}=90\,\text{nm}$, and the gate voltages were set to $V_0=400\,\text{mV}$, $v_0=10\,\text{m/s}$, $T=10\,\mu\text{s}$, $\phi_n=(n-1)\pi/2$, $B_n=0$, $V_{\text{CG}}=150\,\text{mV}$. Fig.~\ref{fig:potential-sim} plots the time variation of the signal wave ($s(t)$), the applied voltage waveform ($V_n(t)$), the shuttling velocity ($v(t)=\dot{x}_{\text{QD}}(t)$), and the quantum dot size ($a_x(t)$) from the top panel to the bottom panel, representing the results for $\beta=0,0.5,1$ from left to right. Here, the position of the quantum dot $x_{\text{QD}}$ represents the minimum point of $V_{\text{QD}}$ with respect to $x$. The quantum dot size $a_x$ is given by $a_x=\sqrt{\frac{1}{m_x \omega_x}}$ using the parameter $\omega_x=\sqrt{\frac{\phi''({\bm{r}}_{\text{QD}})}{m_x}}$ found from the curvature at the minimum point $\phi''({\bm{r}}_{\text{QD}})=\frac{\partial^2 \phi}{\partial x^2}|_{{\bm{r}}={\bm{r}}_{\text{QD}}}$. Here, $m_x$ represents the effective mass of the electron in the shuttling direction, which is given by $m_x\simeq 0.19\, m_0$ using the mass of an electron in vacuum $m_0$. As a result of the numerical calculation, as introduced in Sec.~\ref{sec:Implementation}, it was confirmed that \textit{smooth velocity shuttling}, that is, a continuous shuttling velocity as given by Eq.~(\ref{eq:v}), can be realized by frequency-modulating the gate voltage. It is also worth noting that the quantum dot size is kept almost constant over time. In other words, it can be seen that stable shuttling control is achieved even when applying the modulated voltages proposed in this study. Note that the baseline quantum dot size can be flexibly tuned by adjusting constant parameters, such as the gate width $W$, the voltage amplitude $V_0$ and so on.

\subsection{Shuttling fidelity with valley landscape}
\label{sec:fidelity simulation}
\begin{figure*}[t]
\centering
\includegraphics[width=17cm]{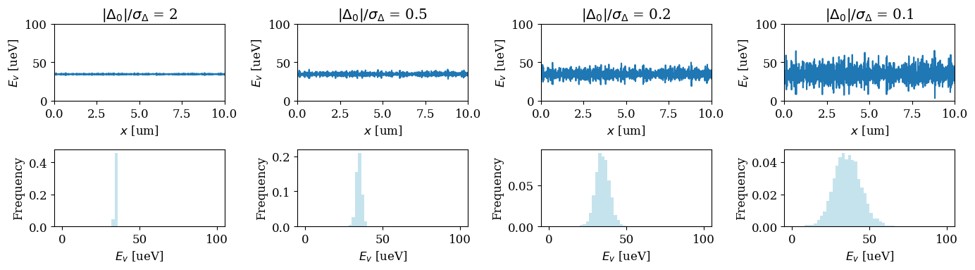}
\caption{Examples of the valley coupling landscape $\Delta(x)$ along the shuttling path. Random spatial variations of the valley coupling are shown for various disorder strengths characterized by the ratio $|\Delta_0|/\sigma_\Delta$. A smaller ratio corresponds to a stronger influence of disorder induced by alloy fluctuations.}
\label{fig:valley}
\end{figure*}

\begin{figure*}[t]
\centering
\includegraphics[width=17cm]{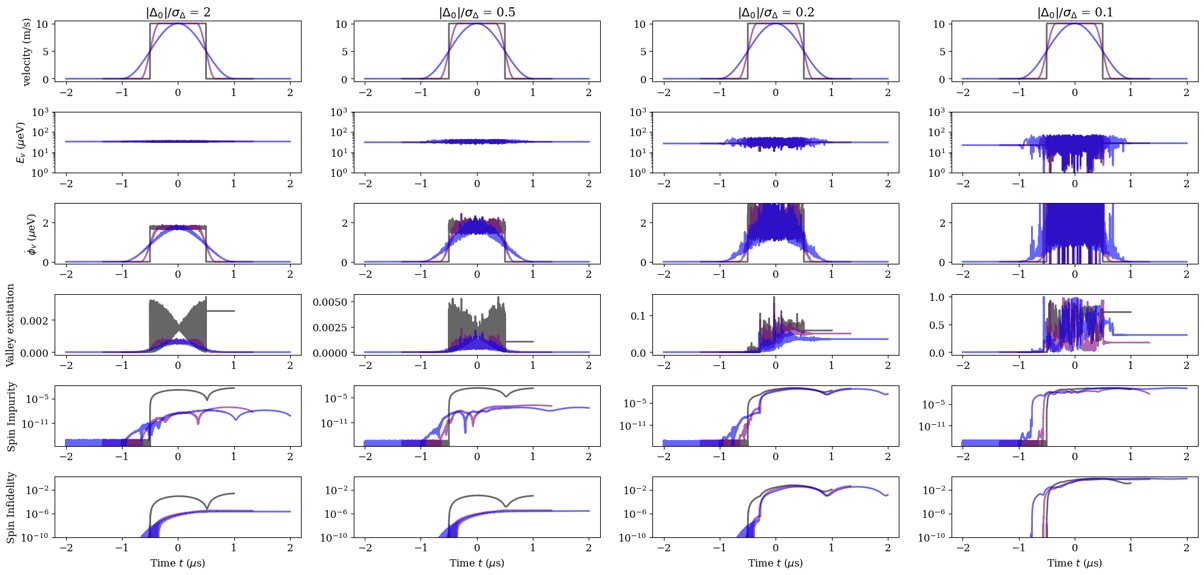}
\caption{Time evolution of spin impurity and spin infidelity during shuttling for a specific fixed valley coupling landscape. Each panel shows the dynamics for a different disorder strength $|\Delta_0|/\sigma_\Delta$. The black, purple, and blue curves correspond to the velocity shaping parameters $\beta=0$ (conventional rectangular window), $\beta=0.5$, and $\beta=1.0$, respectively.}
\label{fig:fidelity}
\end{figure*}

\begin{figure*}[t]
\centering
\includegraphics[width=17cm]{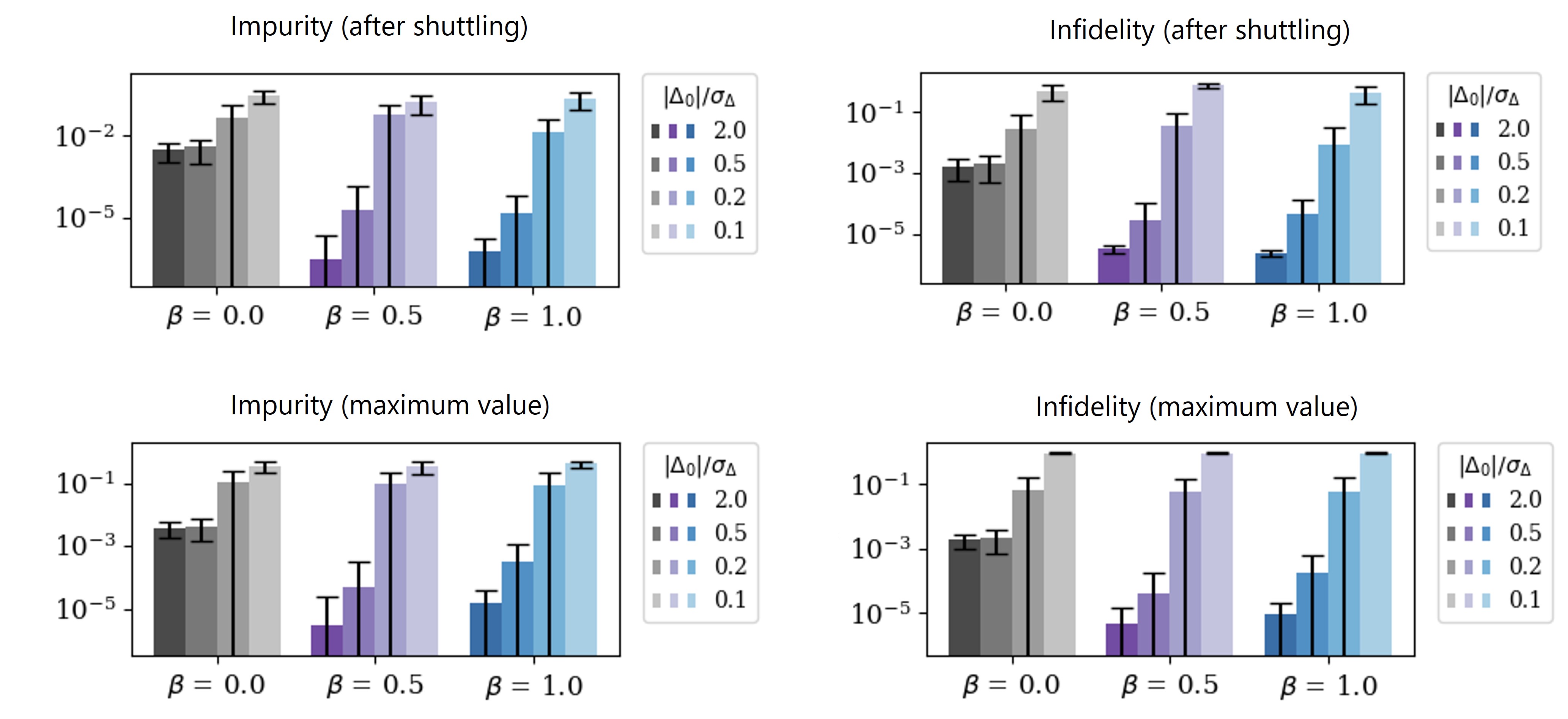}
\caption{Statistical evaluation of spin impurity and infidelity for 50 randomly generated patterns of valley coupling landscapes. 
The random valley landscapes used for the statistical evaluation are generated in the same manner as described in Figs.~\ref{fig:valley} and \ref{fig:fidelity}.
The bar graphs show the mean values and standard deviations (error bars) of spin impurity and spin infidelity at the end of shuttling and these maximum values. Evaluations are performed across different velocity shaping parameters $\beta$ (distinguished by color) and various disorder strengths $|\Delta_0|/\sigma_\Delta$.}
\label{fig:statistic}
\end{figure*}

We mentioned in Sec.~\ref{sec:Proposed Method} that the proposed method, \textit{smooth velocity shuttling}, can lead to the suppression of valley excited state transitions during shuttling. However, in Sec.~\ref{sec:Proposed Method}, sufficient considerations were not given to (i) the relationship between valley excitations and shuttling fidelity, and (ii) the spatial non-uniformity of valley coupling. Therefore, in this section, we extend the physical model to analyze (i) and (ii), and numerically verify the improvement effect of shuttling fidelity by the proposed method.

\subsubsection{Shuttling fidelity}
First, we extend the model to evaluate the shuttling fidelity. Specifically, we extend the model to include not only the valley but also the spin degrees of freedom. Hamiltonians for the extended valley-spin system have been introduced in previous literatures \cite{2024-David-et-al,2026-Romero-et-al}, and we consider a similar model in this study. Specifically, the Hamiltonian of the stationary system is given in the form:
\begin{align}
H_0\,=\,
H_{v0}\otimes I
+
I\otimes H_s
+
H_{v0,s}
\,,
\label{eq:H_0}
\end{align}
Here, $H_{v0}$ is given by Eq.~(\ref{eq:H_v0}). $H_s$ is the Hamiltonian of the spin two-level system ${\ket{\uparrow},\ket{\downarrow}}$, and below we set
\begin{align}
H_s\,=\,
\frac{1}{2}
\begin{pmatrix}
E_Z & 0 \\
0 & -E_Z
\end{pmatrix}
\,=\,
\frac{E_Z}{2}\sigma_z
\,,
\end{align}
where $E_Z$ is the Zeeman splitting, and $\sigma_z$ is the Pauli matrix defined in the space spanned by the spin two-level system ${\ket{\uparrow},\ket{\downarrow}}$ in the stationary system, that is, defined by $\sigma_z=\ket{\uparrow}\!\bra{\uparrow}-\ket{\downarrow}\!\bra{\downarrow}$. $H_{v0,s}$ is the Hamiltonian describing the valley-spin mixing, given by
\begin{align}
H_{v0,s}
\,=\,
-\kappa_z
\begin{pmatrix}
0 & e^{i\phi_v} \\
e^{-i\phi_v} & 0
\end{pmatrix}
\otimes
\begin{pmatrix}
1 & 0 \\
0 & -1
\end{pmatrix}
\,.
\end{align}
$\kappa_z$ is a parameter characterizing the strength of valley-spin mixing, and is typically $\kappa_z\simeq\mathcal{O}(10^{-3}\,\mu\text{eV})$ \cite{2024-David-et-al,2026-Romero-et-al}.
The minus sign is just the convention used in this work.

In the consideration of shuttling fidelity, it is clear to perform a basis transformation to a rotating coordinate system based on the instantaneous eigenstates of the valley. When the Hamiltonian of the stationary system (\ref{eq:H_0}) is transformed to the rotating coordinate system based on the instantaneous eigenstates of the valley, the effective Hamiltonian after the transformation becomes as follows:
\begin{align}
H\,=\,
H_{v}\otimes I
+
I\otimes H_s
+
H_{v,s}
\,.
\label{eq:H}
\end{align}
$H_{v}$ is given by Eq.~(\ref{eq:H_v}). $H_{v,s}$ becomes
\begin{align}
H_{v,s}
\,=\,
-\kappa_z
\begin{pmatrix}
1 & 0 \\
0 & -1
\end{pmatrix}
\otimes
\begin{pmatrix}
1 & 0 \\
0 & -1
\end{pmatrix}
=
-\kappa_z\,
\tilde{\tau}({\bm{r}}_{\text{QD}})
\otimes
\sigma_z
\,,
\label{eq:H_vs}
\end{align}
Equation (\ref{eq:H_vs}) indicates that due to valley-spin mixing, the effective value of the Zeeman splitting differs between the valley ground state and the excited state \cite{2024-Cifuentes-et-al}.

When valley-spin mixing exists, if the valley state changes during shuttling, the spin state after shuttling (the reduced density matrix obtained by tracing out the valley degree of freedom) generally becomes a mixed state. Such changes in the spin state should be avoided as much as possible. Below, we introduce two metrics to evaluate such spin state changes that should be avoided.

\paragraph*{Spin purity} 
Let the quantum state at time $t$ be $\rho_{vs}(t)$. $\rho_{vs}(t)$ is a state defined in the composite system of valley and spin, and its time evolution is given by the time-dependent Hamiltonian (\ref{eq:H}). For $\rho_{vs}(t)$, we consider the reduced density matrix obtained by tracing out the valley degree of freedom, $\rho_{s}(t)=\text{Tr}_v[\rho_{vs}(t)]$. This reduced density matrix $\rho_{s}(t)$ represents the spin state at time $t$. For this $\rho_{s}(t)$, we define the purity $P_s(t)=\text{Tr}[\rho_s(t)^2]$. $P_s$ takes a value of $0\leq P_s\leq 1$ and is a quantity characterizing the decay of the norm of the Bloch vector. When $\rho_{s}$ is a pure state, $P_s=1$, and when $\rho_{s}$ is a completely mixed state, $P_s=0$. Ideally, it is desirable that the spin state is maintained as a pure state even after undergoing shuttling, that is, $P_s=1$.

\paragraph*{Spin fidelity}
Since the spin purity $P_s$ only evaluates the decay of the norm of the Bloch vector, it does not include information regarding the direction of the Bloch vector. Therefore, we introduce the spin fidelity ($\mathcal{F}$) as follows. First, we calculate the spin state $\rho_{s}(t)=\text{Tr}_v[\rho_{vs}(t)]$ at time $t$. We compare this Bloch vector with the Bloch vector of the spin state $\rho_{s,\text{ref}}(t)$ in the case where there is no valley transition (assuming $\dot{\phi}_v/E_v=0$). That is, the fidelity between $\rho_s(t)$ and $\rho_{s,\text{ref}}(t)$ is defined as the spin fidelity ($\mathcal{F}$). Specifically, we calculate $\mathcal{F}(t)=(\text{Tr}\sqrt{\rho_{s,{\text{ref}}}(t) \sqrt{\rho_{s}(t)} \rho_{s,{\text{ref}}}(t)})^2$. Ideally, it is desirable that the spin state does not relatively change even after undergoing shuttling, that is, $\mathcal{F}=1$.

\subsubsection{Valley landscape}

Next, we consider the spatial non-uniformity of valley coupling. In Sec.~\ref{sec:Proposed Method}, for simplicity, it was assumed that the valley coupling $\Delta$ is spatially uniform. However, in actual silicon quantum devices, the value of $\Delta$ randomly fluctuates in space due to variations at the silicon interface \cite{2022-Wuetz-et-al,2023-Losert-et-al,2023-Degli-et-al,2024-Volmer-et-al,2025-Volmer-et-al,2025-Woods-et-al, 2026-HRL}. That is, $\Delta$ can be expressed in the form $\Delta=\Delta_0+\Delta_\delta$ where $\Delta_0$ is the deterministic contribution from an ideal interface without variations, and $\Delta_\delta$ represents the randomness originating from interface variations. In this study, we introduce the statistical model from a previous studies \cite{2022-Wuetz-et-al,2023-Losert-et-al,2025-Woods-et-al} as $\Delta_\delta$. Specifically, the real and imaginary parts of $\Delta_\delta$ are obtained by spatially averaging a Gaussian distribution with a mean of 0 and a variance of $\sigma_\Delta/\sqrt{2}$ over a quantum dot of size $a_x$ (a Gaussian distribution with center $x_{\text{QD}}$ and variance $a_x/\sqrt{2}$). $\sigma_\Delta$ is a parameter characterizing the variation of valley coupling. A small $\sigma_\Delta$ corresponds to a situation where the variation of valley coupling is small (``deterministic regime''), and a large $\sigma_\Delta$ corresponds to a situation where the variation of valley coupling is large (``disorder regime'').

Figure \ref{fig:valley} shows examples of the distribution of the absolute value of valley coupling (valley splitting energy). The upper panels show the spatial distribution of $E_v=2|\Delta|$, and the lower panels show the histograms of $E_v$. In this analysis, as $\Delta_0$, we used the value for an ideal interface with an interface tilt (miscut) $\theta$, $\Delta_0=\frac{E_{v0}}{2}e^{-k^2_0\theta^2 a^2_x}e^{-2ik_0\theta x_{\text{QD}}}$ \cite{2007-Friesen-et-al, 2010-Friesen-Coppersmith}, and set $a_x=10\,\text{nm}$, $\theta=0.8^\circ$. Regarding $E_{v0}$, it was set to $E_{v0}=200\,\mu\text{eV}$. Fig.~\ref{fig:valley} shows examples with larger variations in valley coupling from left to right. Specifically, they correspond to the results when choosing the values of $\sigma_\Delta$ as $|\Delta_0|/\sigma_\Delta=2, 0.5, 0.2, 0.1$. As mentioned above, it can be seen that as $\sigma_\Delta$ increases, the variation in valley splitting energy becomes prominent.

\subsubsection{Results}

With the above, we are now ready to evaluate the effects of the proposed method. Below, we show the results of numerically analyzing the time evolution of the valley-spin system (solving time-dependent Schrödinger equation for Hamiltonian (\ref{eq:H})) using {\texttt{QuTiP}} \cite{QuTip}.

Figure \ref{fig:fidelity} shows the numerical calculation results when the spatial distribution of valley coupling and the shuttling time $T=t_f-t_i$ are fixed. From the top panel to the bottom panel of the figure, the time dependence of the shuttling velocity $v$, the valley splitting energy $E_v$, the time derivative of the valley phase $\dot{\phi}_v$, the valley excitation rate $p_v$, the spin impurity $1-P_s$, and the spin infidelity $1-\mathcal{F}$ are shown. The gray, purple, and blue lines in the figure represent the results for $\beta=0,0.5,1$, respectively. From left to right in the figure, the degree of variation in valley coupling is varied. In this analysis, similarly to before, $\Delta_0$ is set to $\Delta_0=\frac{E_{v0}}{2}e^{-k^2_0\theta^2 a^2_x}e^{-2ik_0\theta x_{\text{QD}}}$ with $E_{v0}=200\,\mu{\text{eV}}$, $a_x=10\,\text{nm}$, $\theta=0.8^\circ$. The values of $\sigma_\Delta$ are set to $\sigma_\Delta/|\Delta_0|=2, 0.5, 0.2, 0.1$ from left to right. In addition, regarding the Zeeman splitting and valley-spin mixing, they were set to $E_Z=0.1\,\text{meV}$ and $\kappa_z=10^{-3}\,\mu\text{eV}$. Let us discuss the results. First, focusing on the valley excitation rate $p_v$, it can be confirmed that the maximum value of the valley excitation rate is suppressed by setting $\beta>0$. Regarding the time dependence, while it oscillates violently in the case of $\beta=0$, the change becomes gradual in the case of $\beta>0$. This indicates that non-adiabatic transitions to the excited valley state are stably suppressed when $\beta>0$, which is the behavior expected in Sec.~\ref{sec:Proposed Method}. This suppression of non-adiabatic transitions to the excited valley state also contributes to ensuring the quality of shuttling can be seen from the behaviors of spin impurity and spin infidelity. It can be seen that by setting $\beta>0$, the values of spin impurity and spin infidelity decrease. However, it should be noted that the effect of $\beta>0$ weakens when the variation in valley coupling is large. In the example of Fig.~\ref{fig:fidelity}, while the effect is prominently seen in the case of $|\Delta_0|/\sigma_0\geq 0.5$, the fidelity improvement effect due to $\beta>0$ weakens when $|\Delta_0|/\sigma_0\leq 0.2$.

The analysis in Fig.~\ref{fig:fidelity} was a fidelity evaluation for a certain fixed valley coupling distribution. Therefore, below we generate multiple valley coupling patterns and perform statistical evaluations on spin purity and fidelity. Specifically, we first generate $N_{\text{rand}}=50$ patterns of valley coupling $\Delta({\bm{r}}_{\text{QD}})$. Thereafter, we perform shuttling simulation (Hamiltonian simulations) for each valley landscape pattern to calculate the spin impurity and spin infidelity, evaluating their mean values and variances. 
When we perform the shuttling simulation for the generated valley landscape, the parameter for the shuttling time $T$ is also randomly selected within the range of $10\,\mu\text{s}\,\pm\,0.1\,\mu\text{s}$ to average out the contribution of the oscillation of the valley excitation rate. Fig.~\ref{fig:statistic} shows the calculation results. In this analysis, similarly to before, $\Delta_0$ is set to $\Delta_0=\frac{E_{v0}}{2}e^{-k^2_0\theta^2 a^2_x}e^{-2ik_0\theta x_{\text{QD}}}$ with $E_{v0}=200\,\mu{\text{eV}}$, $a_x=10\,\text{nm}$, $\theta=0.8^\circ$. Also, the Zeeman splitting and valley-spin mixing are set to $E_Z=0.1\,\text{meV}$ and $\kappa_z=10^{-3}\,\mu\text{eV}$ as before. The upper-left panel shows the spin impurity at the shuttling end time, the lower-left shows the maximum value of spin impurity, the upper-right shows the spin infidelity at the shuttling end time, and the lower-right shows the statistical results (mean value and variance). Each figure compares the results when choosing $\beta$ as $\beta=0,0.5,1$, where $\beta=0$ is represented by a gray bar graph, $\beta=0.5$ by a purple bar graph, and $\beta=1$ by a blue bar graph. The difference in the contrast of colors corresponds to the difference in the variation of valley coupling, representing the results for $\sigma_0/|\Delta_0|= 2,0.5,0.2,0.1$ from dark to light.

From these results, it has been clarified that the effectiveness of the proposed {\textit{smooth velocity shuttling}} depends on the relative strength of the disorder. In the regime where disorder is dominant (for example, $|\Delta_0|/\sigma_\Delta \le 0.2$), the local tiny of valley splitting ($|\Delta| \ll |\Delta_0| $) occurs frequently and sharply, so the improvement effect of spin fidelity by smoothing the shuttling velocity profile ($\beta > 0$) was limited. This is because, in such a strongly disordered landscape, non-adiabatic transitions are driven more strongly by the inherent spatial randomness itself rather than by kinematic noise associated with abrupt acceleration and deceleration.

However, in the moderate-to-low disorder regimes, the benefit of implementing a smoothed velocity profile ($\beta \neq 0$) becomes pronounced. 
With respect to the specific dependency on $\beta$, however, care must be taken: in the presence of disorder, simply increasing $\beta$ toward $1.0$ does not monotonically improve the shuttling performance. Indeed, as shown in Fig.~\ref{fig:statistic}, when the disorder strength is around $|\Delta_0|/\sigma_{\Delta} \simeq 0.5$, the window with $\beta = 0.5$ yields a slightly better result in terms of the mean infidelity compared to the fully smoothed $\beta = 1.0$ window. This finding counteracts the simple considerations discussed in Sec.~\ref{sec:Smooth velocity shuttling}, where the effect of disorder was neglected. 
This counterintuitive behavior can be qualitatively understood by considering the trade-off in the active transport duration. Specifically, approaching $\beta = 1.0$ effectively prolongs the time window of active transport [Eq.~(\ref{eq:T-prime})], which presumably increases the electron's exposure to the spatial randomness of the interface. This suggests that while a larger $\beta$ provides better sidelobe suppression, it can simultaneously amplify the susceptibility to phase randomization. Therefore, our results indicate that finding a balance between these two competing factors is necessary in the presence of disorder, as exemplified by the case where the moderate value of $\beta = 0.5$ outperforms the fully smoothed $\beta = 1.0$ profile.

Given that fabrication techniques to suppress interface disorder in recent Si/SiGe devices (for example, Refs~\cite{2023-Losert-et-al,2024-Volmer-et-al}) are continuously improving, our results emphasize that combining such a relatively clean interface with the proposed smooth velocity control holds an important key to achieving high-fidelity spin shuttling in scalable quantum error-correcting architectures.

\section{Summary and Outlook}
In summary, in this paper, we proposed a {\textit{smooth velocity shuttling}} protocol to suppress non-adiabatic valley excitations during electron shuttling in Si/SiGe quantum dots. By adopting frequency modulation based on the Tukey window for gate voltages, we demonstrated that the high-frequency components of the shuttling velocity spectrum can be effectively suppressed. Our statistical numerical simulations revealed that this smooth control significantly reduces spin infidelity in the moderate-to-low disorder regime ($\sigma_\Delta/|\Delta_0| \simeq \mathcal{O}(1)$) of the valley landscape. 

In particular, we emphasized that the time-domain design of the shuttling velocity profile is mathematically equivalent to the design of window functions in signal processing. Recent theoretical studies have proposed methods to find shuttling velocities that suppress non-adiabatic transitions based on numerical optimization \cite{2024-David-et-al, 2024-Oda-et-al, 2026-Romero-et-al}. Complementary to these numerical approaches, our approach provides an analytical perspective that transforms the physical requirement of suppressing valley excitations into the well-known challenge of suppressing high-frequency sidelobes in the spectrum of a window function. This equivalence provides a design guideline for systematically and intuitively constructing optimal velocity profiles without requiring complex optimization processes.

For future work, several research directions remain. While this model primarily focused on the spatial randomness inherent in the valley landscape, a more comprehensive fidelity evaluation needs to incorporate other dynamic decoherence mechanisms, such as charge noise and spin-orbit interaction. Furthermore, extending this concept of smooth velocity control to two-dimensional shuttling trajectories, such as T-junctions and crossbar arrays, will be an indispensable step toward realizing scalable two-dimensional quantum architectures for surface code error correction. Ultimately, the integration of ongoing material-level advancements, such as the reduction of interface disorder and the enhancement of $|\Delta_0|$ via Ge profile engineering, with the control-level smoothing strategy of this method, presents an extremely promising pathway toward achieving the high-fidelity coherent spin transfer required for future fault-tolerant quantum computing.

\begin{acknowledgments}
We thank Yusuke Wachi and Tatsuya Tomaru
for helpful discussions.
This work is supported by JST Moonshot R\&D Grant No. JPMJMS2065 and JPMJMS256H.
\end{acknowledgments}

\bibliography{ref}
\end{document}